\documentclass{PoS}
\usepackage{amsmath}
\usepackage{subfigure}

\title{$\boldsymbol B\to D^\ast\ell\nu$ at non-zero recoil}

\ShortTitle{\boldmath$B\to D^\ast\ell\nu$ at non-zero recoil}

\author{\speaker{Alejandro Vaquero Avil\'es-Casco}\\
        Department of Physics and Astronomy, University of Utah, Salt Lake City, UT 84112-0830, USA\\
        E-mail: \email{alexvaq@physics.utah.edu}}

\author{Carleton DeTar\\
        Department of Physics and Astronomy, University of Utah, Salt Lake City, UT 84112-0830, USA\\
        E-mail: \email{detar@physics.utah.edu}}

\author{Aida X. El-Khadra\\
        Department of Physics, University of Illinois, Urbana, IL 61801-3080, USA and \\
	Fermi National Accelerator Laboratory, Batavia, IL 60510-5011, USA\\
        E-mail: \email{axk@illinois.edu}}

\author{Andreas S. Kronfeld\\
        Fermi National Accelerator Laboratory, Batavia, IL 60510-5011, USA\\
        E-mail: \email{ask@fnal.gov}}

\author{Jack Laiho\\
        Department of Physics, Syracuse University, Syracuse, NY 13244-1130, USA\\
        E-mail: \email{jwlaiho@syr.edu}}

\author{Ruth S. Van de Water\\
        Fermi National Accelerator Laboratory, Batavia, IL 60510-5011, USA\\
        E-mail: \email{ruthv@fnal.gov}}

\author{(Fermilab Lattice and MILC Collaborations)}

\abstract{We present preliminary blinded results from our analysis of the form factors for $B\rightarrow D^\ast\ell\nu$ decay at non-zero recoil. Our analysis includes 15 MILC asqtad ensembles with $N_f=2+1$
          flavors of sea quarks and lattice spacings ranging from $a\approx 0.15$ fm down to $0.045$ fm. The valence light quarks employ the asqtad action, whereas the $b$ and $c$ quarks are treated using the
          Fermilab action. We discuss the impact that our results will have on $\left|V_{cb}\right|$ and $R(D^\ast)$.}

\FullConference{The 36th Annual International Symposium on Lattice Field Theory - LATTICE2018\\
		22-28 July, 2018\\
		Michigan State University, East Lansing, Michigan, USA.}

\begin{document}

\section{Introduction}

During the last few years the CKM~\cite{Cabibbo:1963yz, Kobayashi:1973fv} matrix element $V_{cb}$ has been at the center of a discussion regarding the unitarity triangle and the search for new physics.
According to the latest HFLAV report~\cite{Amhis:2016xyh}, there is a $2\sigma$ tension between the inclusive and the exclusive determinations, as well as a combined $\sim 4\sigma$ tension between the Standard
Model (SM) predictions and experimental measurements in the $R(D)$-$R(D^\ast)$ plane. Recent developments \cite{BIGI2017441, GRINSTEIN2017359} point, however, towards a simple resolution for the first of these
tensions. There is some evidence that the CLN parametrization~\cite{Caprini:1997mu} is not the optimal one, and might be responsible for the inclusive-exclusive discrepancy (for a review on the current
understanding of the tensions, see~\cite{Tanabashi:2018oca}). On the other hand, none of the existing calculations of $R(D^\ast)$~\cite{Fajfer:2012vx,Bernlochner:2017jka,Bigi:2017jbd,Jaiswal:2017rve} comes
from lattice gauge theory, the only first-principles, non-perturbative tool available to tackle QCD. To solve these matters, a calculation of the form factors of the decay at non-zero recoil is urgently needed.
This work aims to address this issue by performing the first\footnote{At this conference another lattice QCD group presented another calculation at an advanced stage, see Ref.~\cite{Kaneko:2018mcr}.} complete
analysis of the $B\rightarrow D^\ast\ell\nu$ at non-zero recoil on the lattice. Here we present a preliminary result for the form factors, whose normalization is blinded by an overall multiplicative factor.

\section{Notation and definitions}

The Standard Model prediction for the differential rate for exclusive $B \to D^* \ell \nu$ decay can be written in terms of the recoil parameter $w = v_{D^\ast}\cdot v_B$,
\begin{equation}
\frac{d\Gamma}{dw}\left(B\rightarrow D^\ast\ell\nu\right) = \frac{G_F M_B^5}{48\pi^2}\left(1-r^2\right)\sqrt{w^2-1}\chi(w)\left|\eta_{\text{EW}}\right|^2\left|V_{cb}\right|^2\left|\mathcal{F}(w)\right|^2,
\label{dRate}
\end{equation}
where $v_X = p_X/m_X$ are the four velocities of the $B$ and $D^\ast$ mesons, $\eta_{\text{EW}}$ is a correction factor that accounts for electroweak effects, $r = M_{D^\ast}/M_B$, $\mathcal{F}(w)$ is a
function that represents the probability amplitude, to be calculated in lattice QCD, and $\chi(w)$ gathers all the remaining kinematic factors. The function $\mathcal{F}$ can be expressed in
terms of the helicity amplitudes $H_{\pm,0}$ as,
\begin{equation}
\chi(w)\left|\mathcal{F}(w)\right|^2 = \frac{1 - 2wr + r^2}{12M_BM_{D^\ast}(1-r)^2}\left(H_0^2(w) + H_+^2(w) + H_-^2(w)\right).
\end{equation}
The helicity amplitudes, in turn, depend on the $h_X(w)$ form factors, motivated by heavy quark effective theory (HQET),
\begin{align}
H_0  (w) =& \frac{\sqrt{M_BM_{D^\ast}}}{1 - 2wr + r^2}(w+1)\left[(w-r)h_{A_1}(w) - (w-1)(rh_{A_2}(w) + h_{A_3}(2))\right], \\
H_\pm(w) =& \sqrt{M_BM_{D^\ast}}(w+1)\left(h_{A_1}(w) \pm \sqrt{\frac{w-1}{w+1}} h_V(w)\right).
\end{align}

The form factors are defined following the standard decomposition of the matrix elements of the $V - A$ weak current that mediates the transition,
\begin{align}
\frac{\left\langle D^\ast(p_{D^\ast}, \epsilon^\nu)\right|\mathcal{V}^\mu\left|B(0)\right\rangle}{2\sqrt{M_BM_{D^\ast}}} =& \frac{1}{2}\epsilon^\ast_\nu\varepsilon^{\mu\nu}_{\sigma\rho}
v_{D^\ast}^\sigma v_B^\rho,\label{vecCor} \\
\frac{\left\langle D^\ast(p_{D^\ast}, \epsilon^\nu)\right|\mathcal{A}^\mu\left|B(0)\right\rangle}{2\sqrt{M_BM_{D^\ast}}} =& \frac{i}{2}\epsilon^\ast_\nu\left[g^{\mu\nu} (1+w)h_{A_1}(w) -
v_B^\nu\left(v_B^\mu h_{A_2}(w) + v_{D^\ast}^\mu h_{A_3}(w)\right)\right].\label{axCor}
\end{align}
In this work we compute the $h_X$ form factors defined in Eqs.~\eqref{vecCor}, \eqref{axCor} for several recoil values and use them to reconstruct the function $\mathcal{F}(w)$ as a function of $w$.

\section{Simulation details}

\begin{figure}[t]
  \centering
  \includegraphics[width=7.0cm,angle=0]{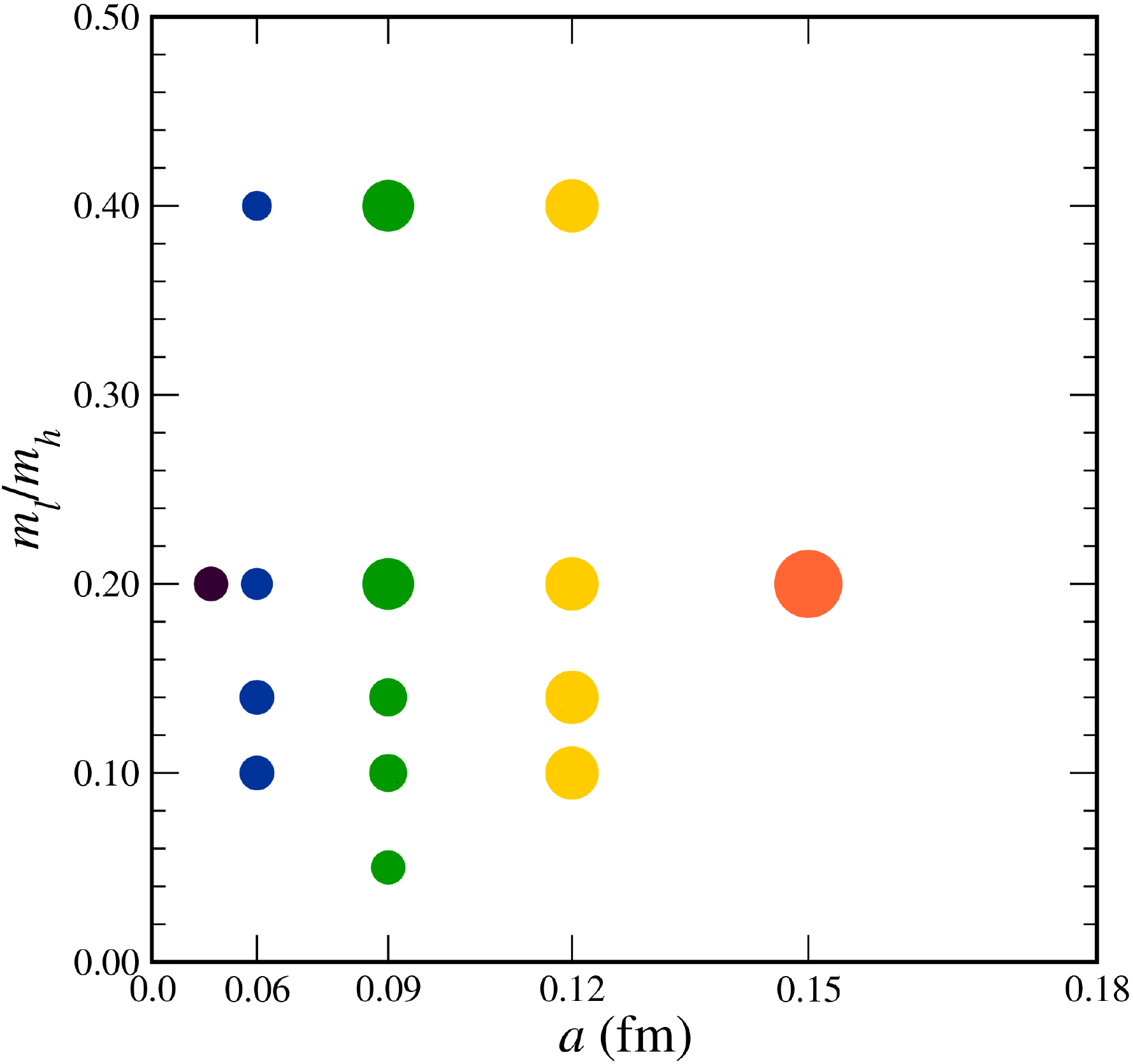}
  \caption{Ensembles used in this calculation. The size of the point gives information about the total statistics available per ensemble, and the vertical axis shows the ratio between the light and the strange
           quark masses. Our smallest pion mass is $M_\pi\approx 180$ MeV.\label{ensFig}}
\end{figure}

For this calculation we employ 15 ensembles of $N_f=2+1$ asqtad~\cite{Lepage:1998vj} sea quarks~\cite{Bazavov:2009bb}. The strange quark is approximately tuned to its physical value, whereas the available
light quark masses and the lattice spacings are shown in Fig.~\ref{ensFig}. The heavy quarks use the clover action with the Fermilab interpretation~\cite{ElKhadra:1996mp}. In our correlators, the $B$ meson is
always at rest, whether the $D^\ast$ meson carries the momentum. Our calculations are done at $\mathbf{p}^2 = 0, \left(2\pi/L\right)^2, \left(4\pi/L\right)^2$ in lattice units, where $L$ is the
spatial size of our lattice. For the non-zero momentum case, we distinguish between the different orientations of the momentum with respect to the polarization of the $D^\ast$ meson $\epsilon^\nu$ and the
current, in order to isolate the form factors in \eqref{vecCor} and \eqref{axCor}.

\section{Lattice results}

For the analysis we largely follow the procedures outlined in our previous works \cite{Lattice:2015rga,Bailey:2014tva,Aviles-Casco:2017nge}. We extract the values of the unrenormalized form factors $h_X(w)$
from the analysis of the two- and three-point functions, then our results are first renormalized and then corrected to adjust the values of the heavy quark masses to their physical value. Blinding is
introduced at the level of the renormalization factors $\rho_{V,A}$\footnote{In this work we use the \emph{mostly non-perturbative renormalization} scheme. The $\rho_{V,A}$ factors mentioned here correspond
to the perturbative component of the renormalization factor for our vector ($V$) and axial ($A$) currents. Our ratios are constructed in such a way that the non-perturbative part cancels out.}: all our
$\rho_{V, A}$ factors are multiplied by an undisclosed random factor close to one. This random factor is known only to one collaboration member who is not working on the analysis. At the present stage of the
analysis, we are still working with blinded data.

\begin{figure}[h]
  \centering
  \subfigure[$h_V(w)$ form factor.]
            {\includegraphics[width=0.48\linewidth,angle=0]{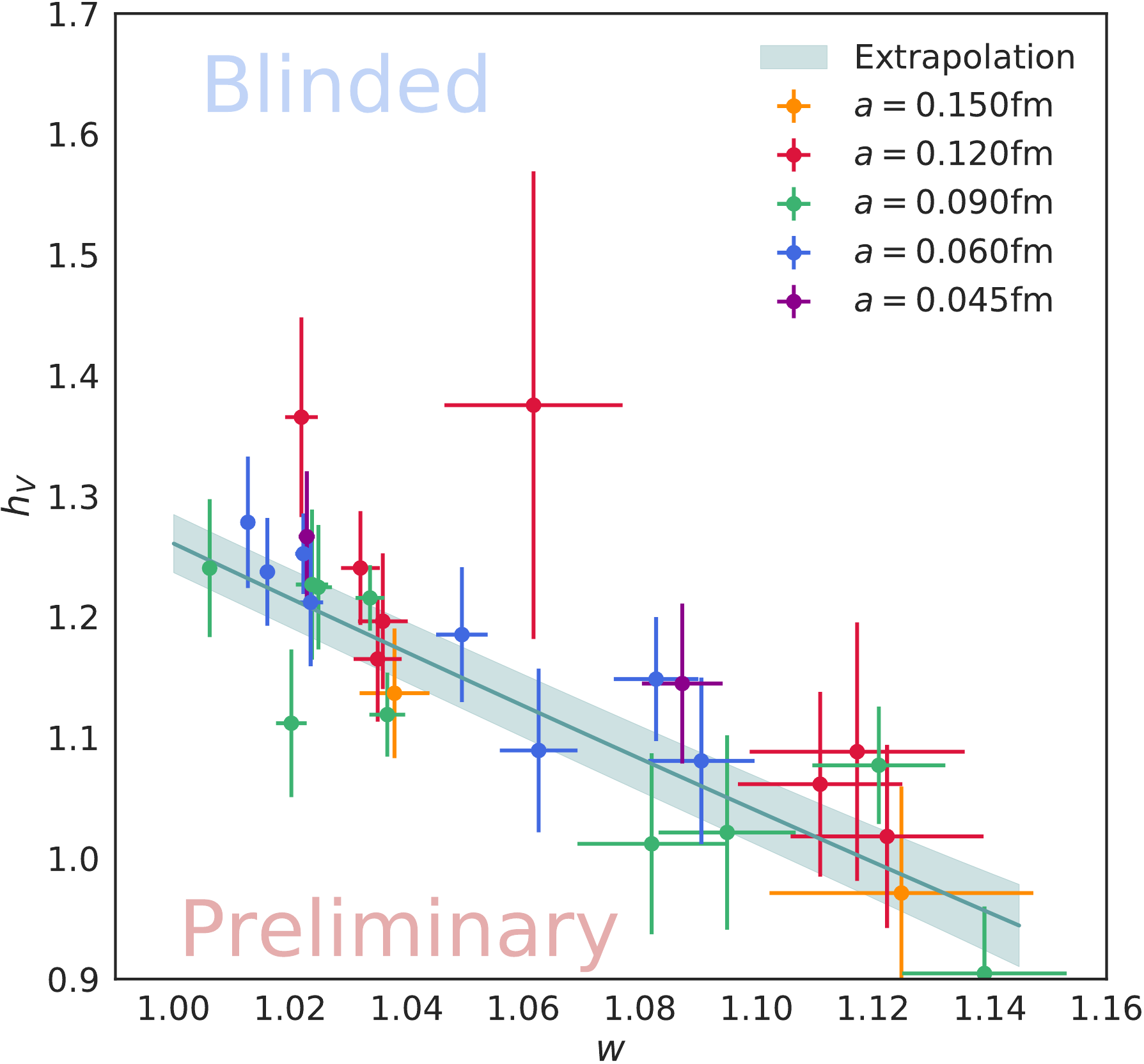} }
  \subfigure[$h_{A_1}(w)$ form factor.]
            {\includegraphics[width=0.48\linewidth,angle=0]{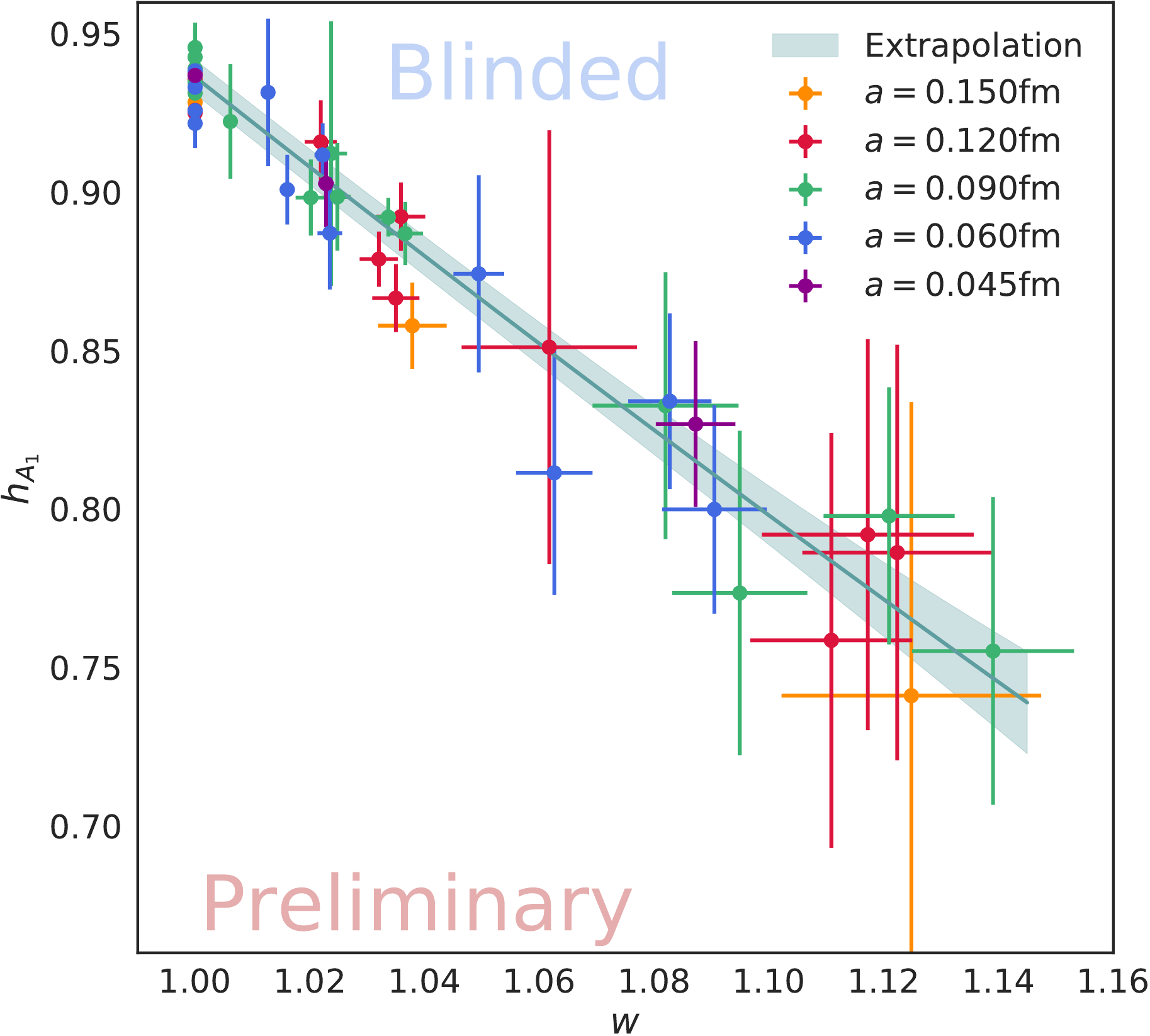} } \\
  \subfigure[$h_{A_2}(w)$ form factor.]
            {\includegraphics[width=0.48\linewidth,angle=0]{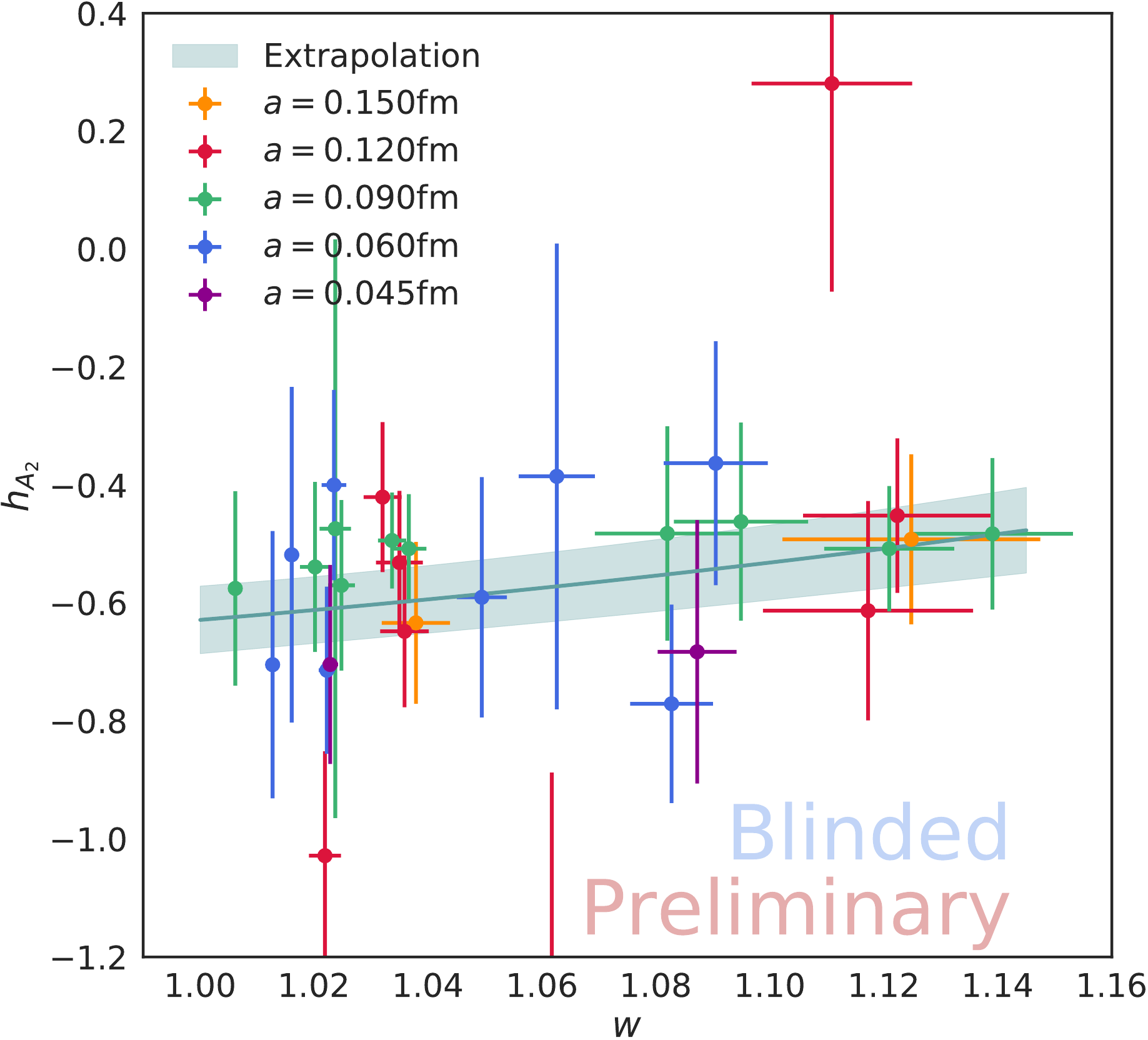} }
  \subfigure[$h_{A_3}(w)$ form factor.]
            {\includegraphics[width=0.48\linewidth,angle=0]{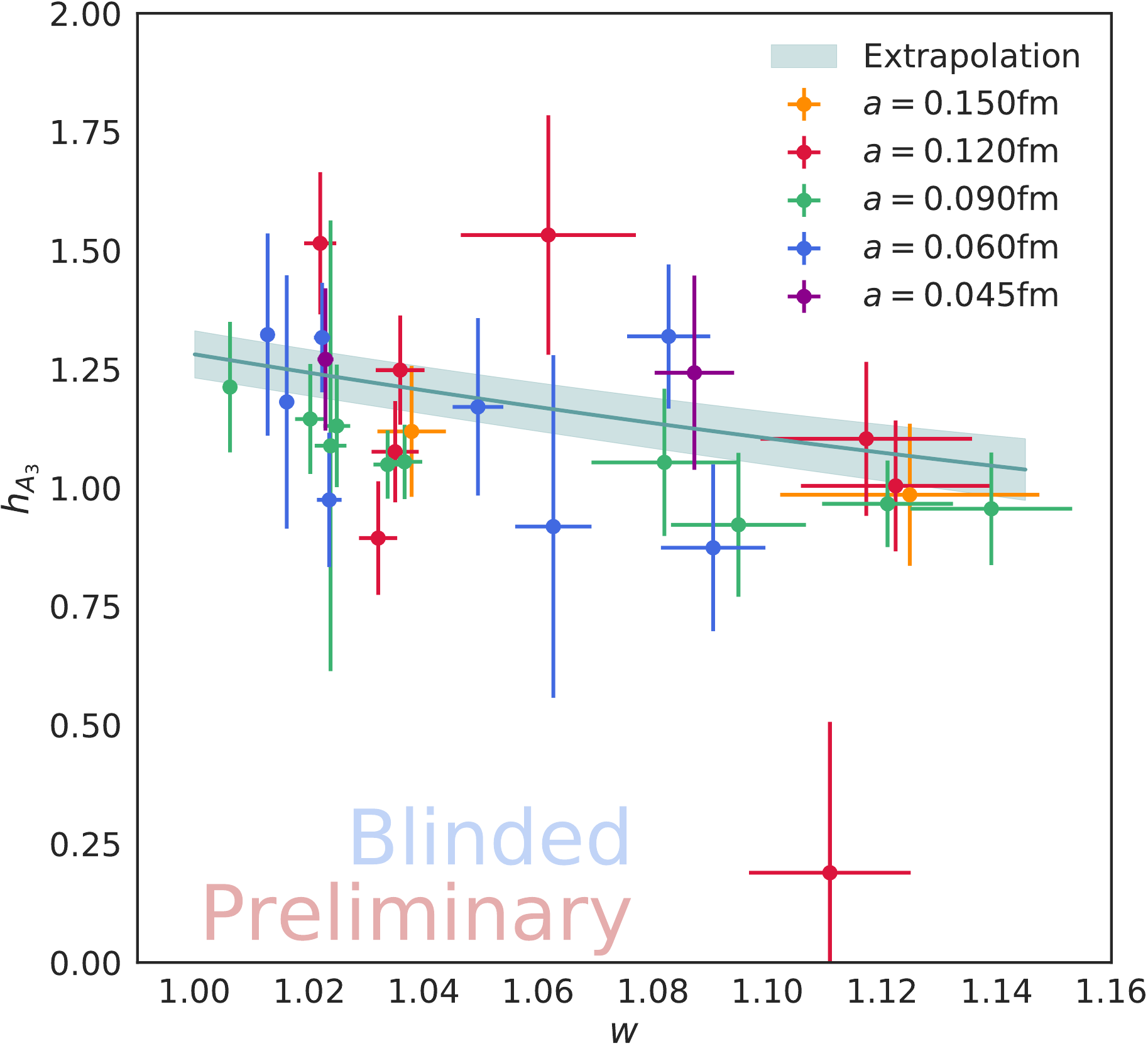} }
  \caption{Preliminary results for $h_V(w)$ and $h_{A_1}(w)$ in the upper row, and $h_{A_2}(w)$ and $h_{A_3}(w)$ in the lower row. The points are the lattice data for different lattice spacings, light quark
           masses and volumes, whereas the band represents the result of the chiral-continuum fit.\label{fFact}}
\end{figure}

Figure~\ref{fFact} gathers all the data for the form factors, after the blinded renormalization factors and the correction to the heavy quark masses have been applied. The chiral-continuum fit is done
following the ansatz
\begin{multline}
h_X = 1 + \frac{X(\Lambda_{\textrm{QCD}})}{m_c} + \frac{g_{D^\ast D\pi}}{48\pi^2f^2_\pi r_1^2}\textrm{logs}_{\textrm{SU(3)}}(w, m_l, m_s, \Lambda_{\textrm{QCD}}) - \rho^2(w - 1) + k(w - 1)^2 \\
    + c_1x_l + c_2x_l^2 + c_{a_1}x_{a^2} + c_{a_2}x_{a^2}^2 + c_{a,m}x_lx_{a^2},
\label{Ch-Ct}
\end{multline}
were $x_l = B_0 m_l / \left(2\pi f_\pi\right)^2$ and $x_{a^2} = a^2 / \left(4\pi f_\pi r_1^2\right)^2$. All the form factors are fitted simultaneously, taking into account all the correlations
among them. There are slight variations depending on the form factor: $h_{A_3}$ and $h_V$ follow exactly Eq.~\eqref{Ch-Ct}, but in $h_{A_1}(1)$ Luke's theorem suggests that the leading HQET term should be
proportional to $1/m_c^2$, and $h_{A_2}$ is not normalized to $1$ at tree level, but to zero. The result of the chiral-continuum fits is used in the $z$ expansion to predict the form of
$\left|\mathcal{F}\right|^2$.

\section{\boldmath$z$ Expansion}

\begin{figure}[t]
  \centering
  \subfigure[$\left|\mathcal{F}\right|^2$ from eq.~\eqref{dRate} computed on the lattice.]
            {\includegraphics[width=0.48\linewidth,angle=0]{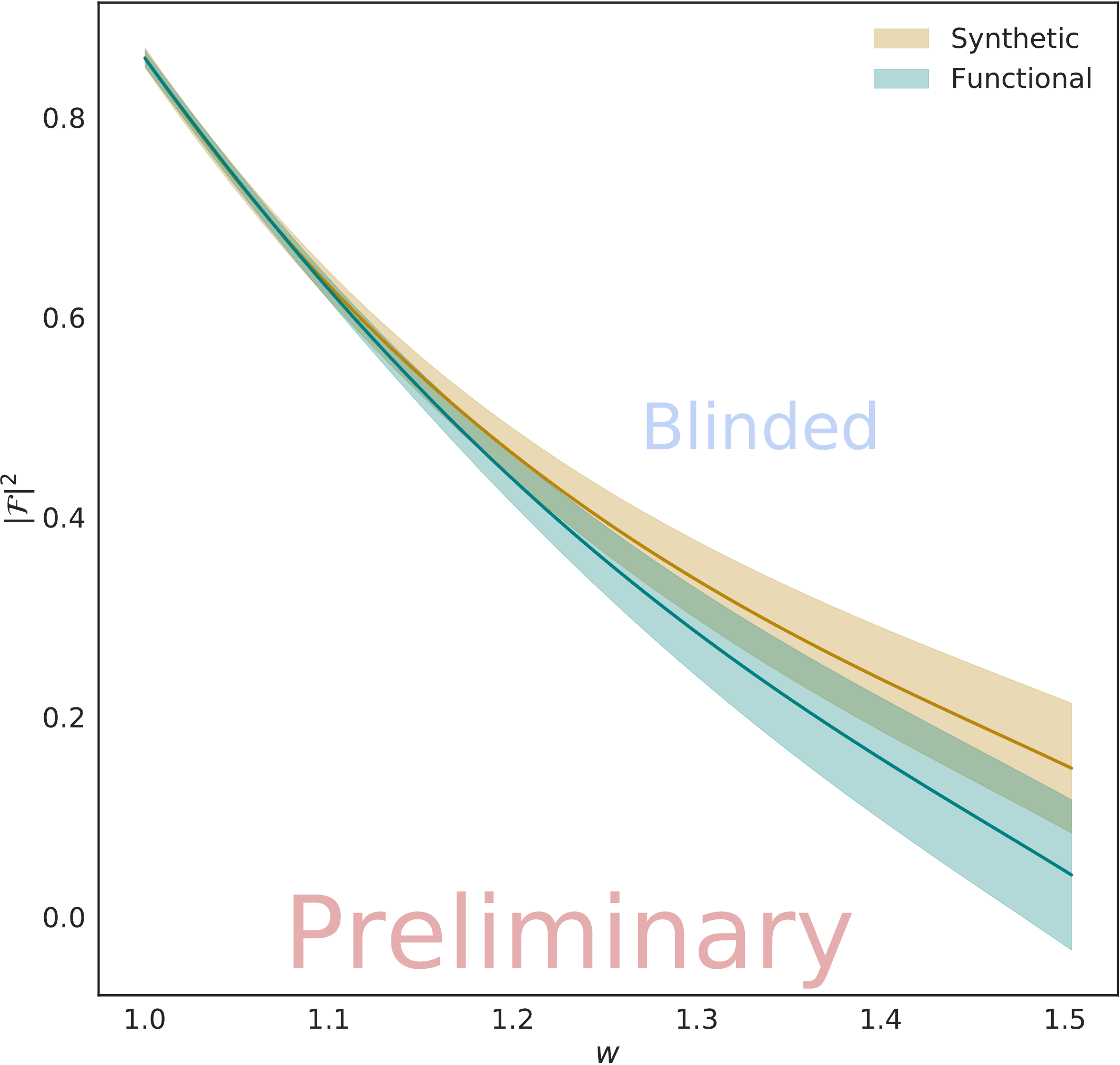}}
  \subfigure[Joint fit using Belle and lattice data.]
            {\includegraphics[width=0.48\linewidth,angle=0]{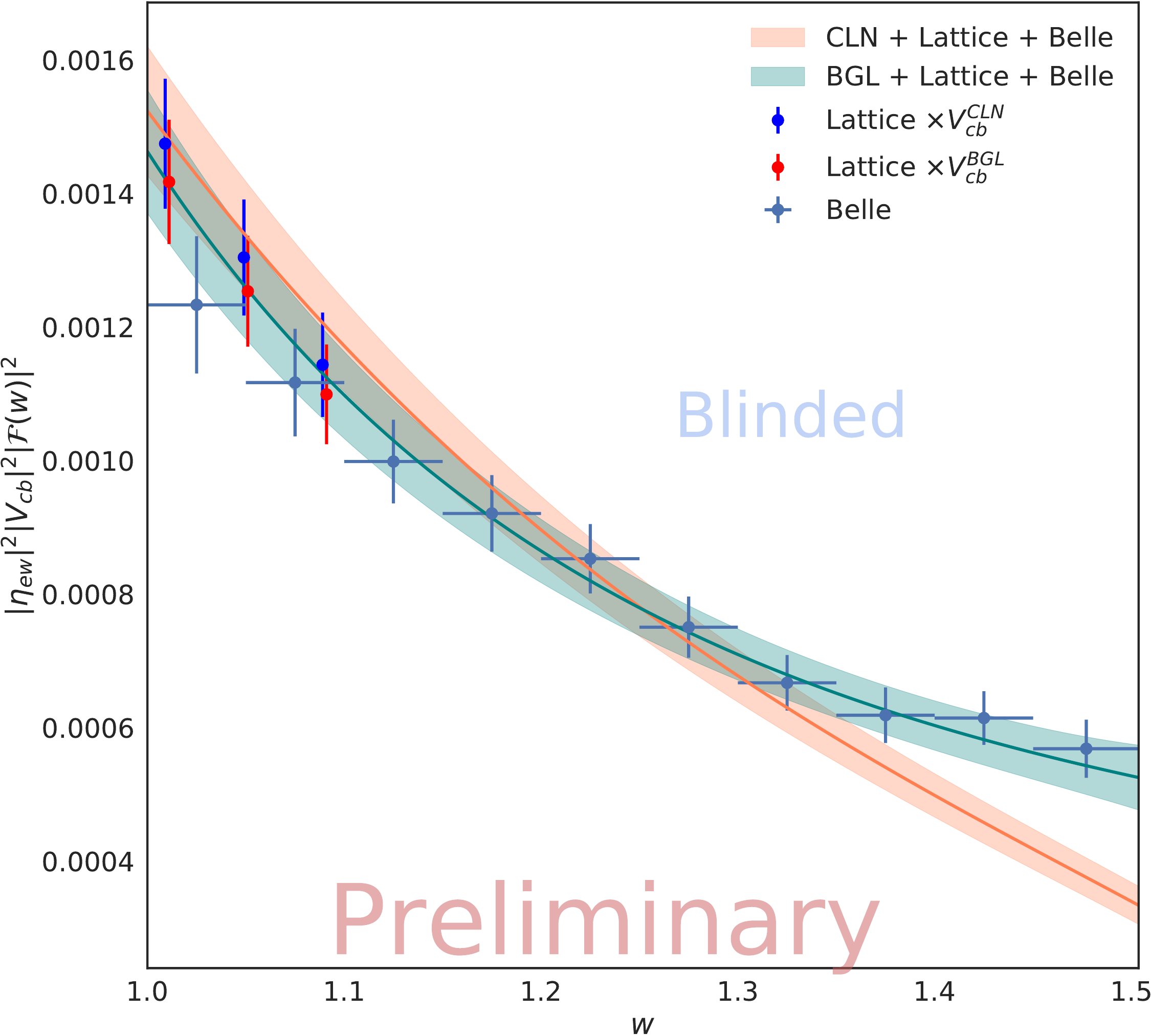}}
  \caption{On the left, pure lattice results for the function $\mathcal{F}$; on the right, joint fit of Belle + lattice data for $\left|V_{cb}\right|^2\left|\eta_{EW}^2\right|\left|\mathcal{F}(w)\right|^2$
           using both the CLN and the BGL parametrizations. $\left|V_{cb}\right|$ is a fit parameter and multiplies the lattice data. For that reason the lattice points for BGL and CLN are slightly different.
           \label{ProbAmp}}
\end{figure}

The fact that $\left|\mathcal{F}(w)\right|^2$ is well known only at zero recoil and that the phase space of the decay vanishes as $\sqrt{w^2 - 1}$ when $w\rightarrow 1$ (see Eq.~\eqref{dRate}) makes an
extrapolation to zero recoil necessary. Even if we can compute the function $\mathcal{F}$ at small recoil (which is the aim of this work), the $z$ expansion provides a model-independent ansatz for a joint fit
with experimental data, involving points at low and high recoil. In our $z$ expansion we use the BGL parametrization~\cite{Boyd:1997kz}, following Refs.~\cite{BIGI2017441, GRINSTEIN2017359}. In particular, we
take the inputs from Ref.~\cite{BIGI2017441}, but we don't see any difference in the final result if the inputs from Ref.~\cite{GRINSTEIN2017359} are used. Since the output of the
chiral-continuum fit described in Eq.~\eqref{Ch-Ct} is a function (and an uncertainty band), we need an extra step in order to generate inputs for the $z$ expansion fit. 
Here we generate synthetic data points from the chiral-continuum fit, where we choose for each form factor three independent points and include the correlations between them. We also explore the functional
method outlined in Ref.~\cite{Lattice:2015tia}. The results for the function $\left|\mathcal{F}\right|^2$ are shown on the left pane of Fig.~\ref{ProbAmp}. This is a purely lattice prediction that doesn't
incorporate any light-cone sum rules (LCSR). We perform a joint fit of synthetic data and experimental data coming from Belle~\cite{Abdesselam:2017kjf}. In this fit we use information only from the
experimental $w$ bins, ignoring the angular distribution. As a test case comparing both parametrizations, we also perform a fit using the CLN parametrization, defined by
\begin{align}
h_{A_1}(w) =& h_{A_1}(1)\left[1 - 8\rho^2 z + \left(53\rho^2 - 15\right)z^2 - \left(231\rho^2 - 91\right)z^3\right] + O(z^4), \label{CLN-hA1}\\
R_1(w)     =& R_1(1) - 0.12(w-1) + 0.05(w-1)^2, \\
R_2(w)     =& R_2(1) + 0.11(w-1) - 0.06(w-1)^2.
\end{align}
The relationship between $R_{1,2}$ and the BGL form factors can be checked in~\cite{BIGI2017441}.
A comparison between our CLN and BGL fits is shown in the right pane of Fig.~\ref{ProbAmp}. The CLN parametrization imposes strict constraints on the behavior of $h_{A_1}(w)$, which seem to be
incompatible with our lattice QCD + Belle data: the high slope at small recoil predicted by lattice QCD and the mild slope determined by experiment at large recoil are difficult to accomodate to
Eq.~\eqref{CLN-hA1}.

\section{Summary and future work}

In this work we show preliminary blinded results for the form factors of the $B\rightarrow D^\ast\ell\nu$ decay at non-zero recoil. While our systematic error analysis is not yet complete, our preliminary
results appear to be in tension with the constraints from the CLN parameterization. We expect that our final results will shed light on the tension between exclusive and inclusive determinations of
$\left|V_{cb}\right|$.

We expect to finalize this analysis and the paper describing it in the coming months. We don't expect that our calculation will yield a $\left|V_{cb}\right|$ determination that is more precise than previous
ones that rely on CLN fits to extrapolate the experimental data to zero recoil. Instead, our calculation will provide new model-independent information on the shape of the form factors at low recoil.

We plan in the coming years to reduce the errors from lattice-QCD, not only for $\left|V_{cb}\right|$, but also for other CKM matrix elements. Our plans include using improved fermionic discretizations for
light and heavy quarks, in order to reduce the chiral, discretization and renormalization errors.

\section{Acknowledgments}

Computations for this work were carried out with resources provided by the USQCD Collaboration, the National Energy Research Scientific Computing Center and the Argonne Leadership
Computing Facility, which are funded by the Office of Science of the U.S. Department of Energy; and with resources provided by the National Institute for Computational Science and the
Texas Advanced Computing Center, which are funded through the National Science Foundation's Teragrid/XSEDE Program. This work was supported in part by the U.S. Department of Energy under
grants No. DE-FC02-06ER41446 (C.D.) and No. DE-SC0015655 (A.X.K.), by the U.S. National Science Foundation under grants PHY10-67881 and PHY14-17805 (J.L.), PHY14-14614 (C.D., A.V.); by
the Fermilab Distinguished Scholars program (A.X.K.); by the German Excellence Initiative and the European Union Seventh Framework Program under grant agreement No. 291763 as well as the
European Union's Marie Curie COFUND program (A.S.K.). Fermilab is operated by Fermi Research Alliance, LLC, under Contract No. DE-AC02-07CH11359 with the United States Department of Energy,
Office of Science, Office of High Energy Physics.

\bibliographystyle{JHEP}
\bibliography{PoSLat18}

\providecommand{\href}[2]{#2}\begingroup\raggedright\begin{thebibliography}{10}

\bibitem{Cabibbo:1963yz}
N.~Cabibbo, \emph{{Unitary symmetry and leptonic decays}},
  \href{https://doi.org/10.1103/PhysRevLett.10.531}{\emph{Phys. Rev. Lett.}
  {\bfseries 10} (1963) 531}.

\bibitem{Kobayashi:1973fv}
M.~Kobayashi and T.~Maskawa, \emph{{CP Violation in the renormalizable theory
  of weak interaction}}, \href{https://doi.org/10.1143/PTP.49.652}{\emph{Prog.
  Theor. Phys.} {\bfseries 49} (1973) 652}.

\bibitem{Amhis:2016xyh}
{\scshape HFLAV} collaboration, \emph{{Averages of $b$-hadron, $c$-hadron, and
  $\tau$-lepton properties as of summer 2016}},
  \href{https://arxiv.org/abs/1612.07233}{{\ttfamily 1612.07233}}.

\bibitem{BIGI2017441}
D.~Bigi, P.~Gambino and S.~Schacht, \emph{{A fresh look at the determination of
  $|V_{cb}|$ from $B\to D^\ast\ell\nu$}},
  \href{https://doi.org/https://doi.org/10.1016/j.physletb.2017.04.022}{\emph{Physics
  Letters B} {\bfseries 769} (2017) 441 }.

\bibitem{GRINSTEIN2017359}
B.~Grinstein and A.~Kobach, \emph{{Model-independent extraction of
  $\left|V_{cb}\right|$ from $\bar{B}\to D^\ast\ell\bar{\nu}$}},
  \href{https://doi.org/https://doi.org/10.1016/j.physletb.2017.05.078}{\emph{Physics
  Letters B} {\bfseries 771} (2017) 359 }.

\bibitem{Caprini:1997mu}
I.~Caprini, L.~Lellouch and M.~Neubert, \emph{{Dispersive bounds on the shape
  of $\bar{B}\to D^\ast\ell\bar{\nu}$ form-factors}},
  \href{https://doi.org/10.1016/S0550-3213(98)00350-2}{\emph{Nucl. Phys.}
  {\bfseries B530} (1998) 153}
  [\href{https://arxiv.org/abs/hep-ph/9712417}{{\ttfamily hep-ph/9712417}}].

\bibitem{Tanabashi:2018oca}
{\scshape Particle Data Group} collaboration, \emph{{Review of particle
  physics}}, \href{https://doi.org/10.1103/PhysRevD.98.030001}{\emph{Phys.
  Rev.} {\bfseries D98} (2018) 030001}.

\bibitem{Fajfer:2012vx}
S.~Fajfer, J.~F. Kamenik and I.~Nisandzic, \emph{{On the $B \to D^* \tau \bar
  \nu_{\tau}$ Sensitivity to New Physics}},
  \href{https://doi.org/10.1103/PhysRevD.85.094025}{\emph{Phys. Rev.}
  {\bfseries D85} (2012) 094025}
  [\href{https://arxiv.org/abs/1203.2654}{{\ttfamily 1203.2654}}].

\bibitem{Bernlochner:2017jka}
F.~U. Bernlochner, Z.~Ligeti, M.~Papucci and D.~J. Robinson, \emph{{Combined
  analysis of semileptonic $B$ decays to $D$ and $D^*$: $R(D^{(*)})$,
  $|V_{cb}|$, and new physics}},
  \href{https://doi.org/10.1103/PhysRevD.95.115008,
  10.1103/PhysRevD.97.059902}{\emph{Phys. Rev.} {\bfseries D95} (2017) 115008}
  [\href{https://arxiv.org/abs/1703.05330}{{\ttfamily 1703.05330}}].

\bibitem{Bigi:2017jbd}
D.~Bigi, P.~Gambino and S.~Schacht, \emph{{$R(D^*)$, $|V_{cb}|$, and the Heavy
  Quark Symmetry relations between form factors}},
  \href{https://doi.org/10.1007/JHEP11(2017)061}{\emph{JHEP} {\bfseries 11}
  (2017) 061} [\href{https://arxiv.org/abs/1707.09509}{{\ttfamily
  1707.09509}}].

\bibitem{Jaiswal:2017rve}
S.~Jaiswal, S.~Nandi and S.~K. Patra, \emph{{Extraction of $|V_{cb}|$ from
  $B\to D^{(*)}\ell\nu_\ell$ and the Standard Model predictions of
  $R(D^{(*)})$}}, \href{https://doi.org/10.1007/JHEP12(2017)060}{\emph{JHEP}
  {\bfseries 12} (2017) 060}
  [\href{https://arxiv.org/abs/1707.09977}{{\ttfamily 1707.09977}}].

\bibitem{Kaneko:2018mcr}
{\scshape JLQCD} collaboration, \emph{{$B \to D^{(*)}\ell\nu$ form factors from
  $N_f\!=\!2+1$ QCD with M\"obius domain-wall quarks}},  vol.~LATTICE2018,
  p.~311, 2018, \href{https://arxiv.org/abs/1811.00794}{{\ttfamily
  1811.00794}}.

\bibitem{Lepage:1998vj}
G.~P. Lepage, \emph{{Flavor symmetry restoration and Symanzik improvement for
  staggered quarks}},
  \href{https://doi.org/10.1103/PhysRevD.59.074502}{\emph{Phys. Rev.}
  {\bfseries D59} (1999) 074502}
  [\href{https://arxiv.org/abs/hep-lat/9809157}{{\ttfamily hep-lat/9809157}}].

\bibitem{Bazavov:2009bb}
{\scshape Fermilab Lattice, MILC} collaboration, \emph{{Nonperturbative QCD
  simulations with 2+1 flavors of improved staggered quarks}},
  \href{https://doi.org/10.1103/RevModPhys.82.1349}{\emph{Rev. Mod. Phys.}
  {\bfseries 82} (2010) 1349}
  [\href{https://arxiv.org/abs/0903.3598}{{\ttfamily 0903.3598}}].

\bibitem{ElKhadra:1996mp}
A.~X. El-Khadra, A.~S. Kronfeld and P.~B. Mackenzie, \emph{{Massive fermions in
  lattice gauge theory}},
  \href{https://doi.org/10.1103/PhysRevD.55.3933}{\emph{Phys. Rev.} {\bfseries
  D55} (1997) 3933} [\href{https://arxiv.org/abs/hep-lat/9604004}{{\ttfamily
  hep-lat/9604004}}].

\bibitem{Lattice:2015rga}
{\scshape Fermilab Lattice, MILC} collaboration, \emph{{$B\to D\ell\nu$ form
  factors at nonzero recoil and $\left|V_{cb}\right|$ from 2+1-flavor lattice
  QCD}}, \href{https://doi.org/10.1103/PhysRevD.92.034506}{\emph{Phys. Rev.}
  {\bfseries D92} (2015) 034506}
  [\href{https://arxiv.org/abs/1503.07237}{{\ttfamily 1503.07237}}].

\bibitem{Bailey:2014tva}
{\scshape Fermilab Lattice, MILC} collaboration, \emph{{Update of $|V_{cb}|$
  from the $\bar{B}\to D^*\ell\bar{\nu}$ form factor at zero recoil with
  three-flavor lattice QCD}},
  \href{https://doi.org/10.1103/PhysRevD.89.114504}{\emph{Phys. Rev.}
  {\bfseries D89} (2014) 114504}
  [\href{https://arxiv.org/abs/1403.0635}{{\ttfamily 1403.0635}}].

\bibitem{Aviles-Casco:2017nge}
A.~Vaquero Avil\'es-Casco, C.~DeTar, D.~Du, A.~El-Khadra, A.~S. Kronfeld,
  J.~Laiho et~al., \emph{{$\overline{B}\rightarrow D^\ast\ell\overline{\nu}$ at
  non-zero recoil}},
  \href{https://doi.org/10.1051/epjconf/201817513003}{\emph{EPJ Web Conf.}
  {\bfseries 175} (2018) 13003}
  [\href{https://arxiv.org/abs/1710.09817}{{\ttfamily 1710.09817}}].

\bibitem{Boyd:1997kz}
C.~G. Boyd, B.~Grinstein and R.~F. Lebed, \emph{{Precision corrections to
  dispersive bounds on form-factors}},
  \href{https://doi.org/10.1103/PhysRevD.56.6895}{\emph{Phys. Rev.} {\bfseries
  D56} (1997) 6895} [\href{https://arxiv.org/abs/hep-ph/9705252}{{\ttfamily
  hep-ph/9705252}}].

\bibitem{Lattice:2015tia}
{\scshape Fermilab Lattice, MILC} collaboration, \emph{{$|V_{ub}|$ from
  $B\to\pi\ell\nu$ decays and (2+1)-flavor lattice QCD}},
  \href{https://doi.org/10.1103/PhysRevD.92.014024}{\emph{Phys. Rev.}
  {\bfseries D92} (2015) 014024}
  [\href{https://arxiv.org/abs/1503.07839}{{\ttfamily 1503.07839}}].

\bibitem{Abdesselam:2017kjf}
{\scshape Belle} collaboration, \emph{{Precise determination of the CKM matrix
  element $\left| V_{cb}\right|$ with $\bar B^0 \to D^{*\,+} \, \ell^- \, \bar
  \nu_\ell$ decays with hadronic tagging at Belle}},
  \href{https://arxiv.org/abs/1702.01521}{{\ttfamily 1702.01521}}.

\end{thebibliography}\endgroup

\end{document}